\def\year{2024}\relax
\documentclass[letterpaper]{article} 
\usepackage{aaai22}  
\usepackage{times}  
\usepackage{helvet}  
\usepackage{courier}  
\usepackage[hyphens]{url}  
\usepackage{graphicx} 
\usepackage{natbib}  
\usepackage{caption} 
\usepackage{balance}
\DeclareCaptionStyle{ruled}{labelfont=normalfont,labelsep=colon,strut=off} 
\usepackage{algorithm}
\usepackage{algorithmic}
\usepackage{color}
\usepackage{tikz}
\usepackage{xcolor} 
\definecolor{mygray}{RGB}{169,169,169} 
\usetikzlibrary{shapes, shadows} 
\usepackage{enumitem}
\usepackage{lipsum}
\usepackage{adjustbox}

\usepackage{newfloat}
\usepackage{listings}
\floatstyle{ruled}
\newfloat{listing}{tb}{lst}{}
\floatname{listing}{Listing}
\title{Hidden in Plain Sight: Exploring the Intersections of Mental Health, \\ Eating Disorders, and Content Moderation on TikTok}
\author{
    Charles Bickham\equalcontrib\textsuperscript{\rm 1,\rm 2},
    Kia Kazemi-Nia\equalcontrib\textsuperscript{\rm 1,\rm 2, \rm 3},
    Luca Luceri\textsuperscript{\rm 2},
    Kristina Lerman\textsuperscript{\rm 1, \rm 2},
    Emilio Ferrara\textsuperscript{\rm 1,\rm 2}
}
\affiliations{
 
\textsuperscript{\rm 1} Department of Computer Science, Viterbi School of Engineering, University of Southern California, Los Angeles, CA, USA

\textsuperscript{\rm 2}Information Sciences Institute, University of Southern California, Marina Del Rey, CA, USA

\textsuperscript{\rm 3}\small Department of Psychology, Dornsife College of Letters, Arts and Sciences, University of Southern California, Los Angeles, CA, USA 

cbickham@usc.edu, kazemini@usc.edu, lluceri@isi.edu, lerman@isi.edu, emiliofe@usc.edu

}

\usepackage{bibentry}

\begin{document}

\maketitle

\begin{abstract}

Social media platforms actively moderate content glorifying harmful behaviors like eating disorders, which include anorexia and bulimia. However, users have adapted to evade moderation by using coded hashtags. Our study investigates the prevalence of moderation evaders on the popular social media platform TikTok and contrasts their use and emotional valence with mainstream hashtags. We notice that moderation evaders and mainstream hashtags appear together, indicating that vulnerable users might inadvertently encounter harmful content even when searching for mainstream terms.
Additionally, through an analysis of emotional expressions in video descriptions and comments, we find that mainstream hashtags generally promote positive engagement, while moderation evaders evoke a wider range of emotions, including heightened negativity. These findings provide valuable insights for content creators, platform moderation efforts, and interventions aimed at
cultivating a supportive online environment for discussions
on mental health and eating disorders.
\end{abstract}

Warning: This paper discusses eating disorders, which some may find distressing.

\section{Introduction}
This study explores the interplay between mental health and social media, using a case study of TikTok and eating disorders. 
Eating disorders represent a complex mental health condition characterized by disruptions in eating and related behaviors.  
These conditions,  which include anorexia, bulimia, and binge eating disorder (BED)~\cite{walsh2020eating}, have alarming mental health impacts: one study reported that up to 23\% of individuals diagnosed with BED in the US had attempted suicide, with 94\% experiencing a lifetime of mental health problems \cite{keski2021epidemiology}.
The COVID-19 pandemic further complicated the landscape of eating disorders \cite{hogue2019effects,rodgers2020impact,schlegl2020eating}. During this period, there was a notable surge in symptomatology, accompanied by diminished access to treatment for individuals struggling with eating disorders \cite{cooper2022eating}. This crisis highlighted the urgency of understanding the dynamic interplay between external stressors, mental health, and the manifestation of eating disorders.

TikTok, a platform for sharing short-form video content, has billions of users worldwide ~\cite{iqbal2021tiktok} and is especially popular among adolescents. While TikTok provides a creative outlet for many young people, it has also become a space for discussions about mental health and eating disorders \cite{herrick2021just,mccashin2023using}. The brevity and immediacy of TikTok videos, often accompanied by succinct descriptions containing hashtags, 
present an opportunity to investigate how individuals engage with emotional issues, as well as broader questions about the role of social media in mental wellbeing, particularly in adolescents~\cite{frieiro2021social,pruccoli2022use,sha2021research}.

This work explores the emotional landscape of TikTok videos related to eating disorders, focusing on the role of hashtags in the discoverability and organization of content. 
While TikTok heavily moderates searches for harmful content that promotes or glorifies eating disorders \cite{casilli2013online}, users have adapted to  evade moderation through the use of coded hashtags \cite{herrick2021just,cobb2017not}. A study of TikTok by the Center for Countering Digital Hate (CCDH)\footnote{\url{https://counterhate.com/wp-content/uploads/2022/12/CCDH-Deadly-by-Design_120922.pdf#page=44.12}\label{note1}} found content promoting eating disorders relied on coded hashtags to avoid getting flagged by content moderation algorithms. These coded hashtags were based on misspellings, abbreviations, or references to the artist Ed Sheeran, whose name coincidentally begins with ``ED" -- an abbreviation commonly associated with eating disorders.
For the purpose of this study, we refer to such hashtags as \textit{moderation evaders}. In this paper, we analyze the emotional expressions in content across different hashtags pertaining to eating disorders, including moderation evaders and mainstream hashtags. 


\begin{itemize}
    \item \textbf{RQ1:} Do moderation evaders co-occur with mainstream hashtags related to mental health and healthy living?
    
    
    
    \item \textbf{RQ2:} How do emotional expressions in TikTok video descriptions differ between content associated with mainstream hashtags and moderation evaders?
    \item \textbf{RQ3:} How do emotional expressions in TikTok video descriptions correlate with those in user comments, and how does this interaction differ across mainstream hashtags and moderation evaders?
    
\end{itemize}
Our study of the emotional expressions within video descriptions and user comments reveals systematic differences between mainstream hashtags and moderation evaders. While mainstream hashtags tend to promote an overall positive engagement, moderation evaders evoke a broader range of emotions, including heightened negativity. The disparity in emotional engagement suggests that moderation evading hashtags are used for spreading problematic content related to eating disorders. Moreover, since moderation evaders co-occur frequently with mainstream hashtags, this raises concerns about the potential exposure of users to harmful content, as moderation evaders often circumvent platform regulations. 

Our findings underscore the complex emotional landscape of TikTok content related to eating disorders and emphasize the need for tailored moderation strategies and interventions to cultivate a supportive online environment conducive to discussions on mental health and eating disorders. In the subsequent sections, we present related works, our methodology, discuss our results, and draw meaningful conclusions that contribute to the evolving discourse on mental health and eating disorders in the digital age.


\section{Related Works}

\subsection{Social Media and Body Image Concerns}
The rise in social media use has fueled worries about its impact on negative body image concerns. Social comparison has been found to play a negative role in how people think about their body image \cite{hulsing2021triggerwarning,mink2022tiktok,westenberg2023impact,jiotsa2021social,festinger1957social}. Women and young girls especially tend to suffer from negative body image concerns \cite{peng2023negative,liu2022social,hogue2019effects}. Social media platforms that are based on photos are thought of to be more negative when it comes to body image since they tend to be more focused on physical appearance \cite{karsay2021don,rodgers2022social}. With photo-based platforms, there are more opportunities for people, especially women, to self-objectify, internalize appearance ideals, and compare themselves negatively. This can lead to many mental health risks that include but is not limited to unipolar depression, sexual dysfunction, and eating disorders \cite{fredrickson1997objectification}. Additionally, the use of photo-based platforms allows for images to be manipulated. Studies show that there has been a link between lower body image and self-esteem based on posting exposure to manipulated images \cite{chua2016follow,wick2020posting,cohen2018selfie,kleemans2018picture}. Also, many people may not be aware of the images being manipulated, which can lead to a normalization of unrealistic body and beauty ideals~\cite{marks2020pursuit}. Exposure to the thin ideal has been linked to an increase in body dissatisfaction, eating disorder symptoms,
and negative mood in women~\cite{hawkins2004impact, fardouly2015social,tiggemann2020uploading}. The presence of a negative self-body image in an individual is identified as a contributing factor to eating disorders~\cite{manaf2016prevalence}.

\subsection{Social Media and Eating Disorders}
Social media use has been linked with distorted eating over recent years \citet{holland2016systematic,zhang2021relationship}. With the rise of this trend, there have been efforts to prevent eating disorders. \cite{de2022targeting} showed that self-criticism intervention can be a strategy to address this need. Communication strategies have also been looked into to help the prevention of eating disorders \cite{rando2023health}. \citet{lerman2023radicalized} map the relationship between social media and eating disorders to an online radicalization process \cite{kruglanski2014psychology,wang2023identifying}. Exploring the concept of online radicalization within ``pro-ana" communities, \cite{lerman2023radicalized} highlights how social media platforms facilitate radicalized behaviors by creating echo chambers that can normalize distorted eating. The study emphasizes the importance of understanding and quantifying the impact of these online communities to develop strategies aimed at promoting better mental health. \citet{branley2017pro} details how easy it is for someone to ``stumble" upon potentially harmful ED content without explicitly searching for it.

Specifically on TikTok, \citet{herrick2021just} analyzes the mainstream hashtag \#edrecovery, while \cite{greene2023follow} analyzes the textual content, using mixed methods, on the hashtag \#bedrecovery. \cite{hung2022content} conducts content analysis on the hashtags \#fitspiration and \#thinspiration. \citet{greene2023visions} performed a comparative analysis on pro-recovery communities across five eating disorder hashtags: \#anarecovery,\ \#arfidrecovery, \#bedrecovery, \#miarecovery, and \#orthorexiarecovery. \cite{dondzilo2023association} supports indirect connections between involvement with appearance/eating-related content on social media and eating disorder symptoms, with mediation through elevated exposure to recommended content in this category and increased levels of upward social media appearance comparisons.

The impact of social media and eating disorders on the youth have also been studied \cite{lonergan2020protect,pruccoli2022use,corzine2023social,salomon2020examining}. These studies emphasize the need for an understanding of the relationship between social media use and eating disorders among young adults and children.

\begin{table*}[t!]
\centering
\begin{tabular}{|c|c|c|c|c|c|c|c|}
\hline
\textbf{Cluster} & \textbf{Count} & \textbf{Average Likes} & \textbf{Average Shares} & \textbf{Average Plays} & \textbf{Average Comments} \\
\hline
ED recovery & 182 & 239,976.80 & 4,915.86 & 1,948,781.66 & 1,868.81 \\
Healthy living & 80 & 149,775.51 & 3,299.18 & 1,399,869.32 & 1,317.07 \\
Body positivity/acceptance & 31 & 31,887.22 & 853.84 & 353,926.05 & 276.77 \\
Juice-related content & 35 & 22,947.38 & 752.35 & 233,925.57 & 243.05 \\
Beauty and body positivity & 81 & 296,309.77 & 6,105.77 & 2,491,468.19 & 2,327.29 \\
Miscellaneous & 86 & 317,098.94 & 4,465.23 & 2,417,362.08 & 2,637.50 \\
Music promotion & 38 & 25,090.83 & 875.28 & 282,339.19 & 227.03 \\
\hline
\end{tabular}
\caption{Statistics for each hashtag community}
\label{tab:averaged_modularity_stats}
\end{table*}

\section{Methodology \string& Data}
\subsection{Dataset}
For the purpose of this research, we collected information about TikTok videos focusing on content associated with various issues related to body image, dieting, and eating disorders. To collect the data, we curated a set of hashtags that reflect the diversity of topics within the TikTok platform. See the Appendix for the full list of keywords. Spanning the timeframe from December 2016 to April 2023, the dataset has a total of 14,816 posts and 562,856 comments associated with these videos.

The dataset encompasses various aspects of TikTok videos with each entry containing textual information. Textual information is captured through the video description (Description) and a list of hashtags used in the video (Challenges). Additionally, the dataset includes the comments the videos received.


\subsection{Hashtag Co-Occurrence Graph}
A hashtag co-occurrence graph operates as a network that outlines the relationships among hashtags present in a series of social media posts. When two hashtags share an edge, it signifies their joint appearance in a post, and the weight of this edge reflects the frequency of these occurrences—essentially measuring how often they appeared together in our dataset.

\subsection{Measuring Emotions}
Textual content conveys signals related to emotions and feelings, encompassing both positive sentiments, such as joy and love, and negative ones like anger and disgust. In our approach, we employ an emotion detection tool inspired by SpanEmo \cite{alhuzali2021spanemo}, named Demultiplexer (Demux) \cite{chochlakis2023leveraging}. The tool takes the categories (in this case, emotions) as the first input sequence and the actual content as the second sequence. The contextual embeddings specific to each emotion contribute to deriving probabilities for individual emotions. Our application of Demux extends to every video description and comment. The emotions included were anger, anticipation, disgust, fear, joy, love, optimism, pessimism, sadness, surprise, trust, and none. For the purpose of this study, anticipation, pessimism, surprise, and trust were grouped into a single category named ``other emotions" similar to \cite{lerman2023radicalized}.

\section{Results}

The methods detailed above were used to analyze our dataset and answer our research questions about the co-occurrences and emotional patterns across mainstream hashtags and moderation evaders.

\begin{figure*}[t]
    \centering
    \includegraphics[width=0.96\linewidth]{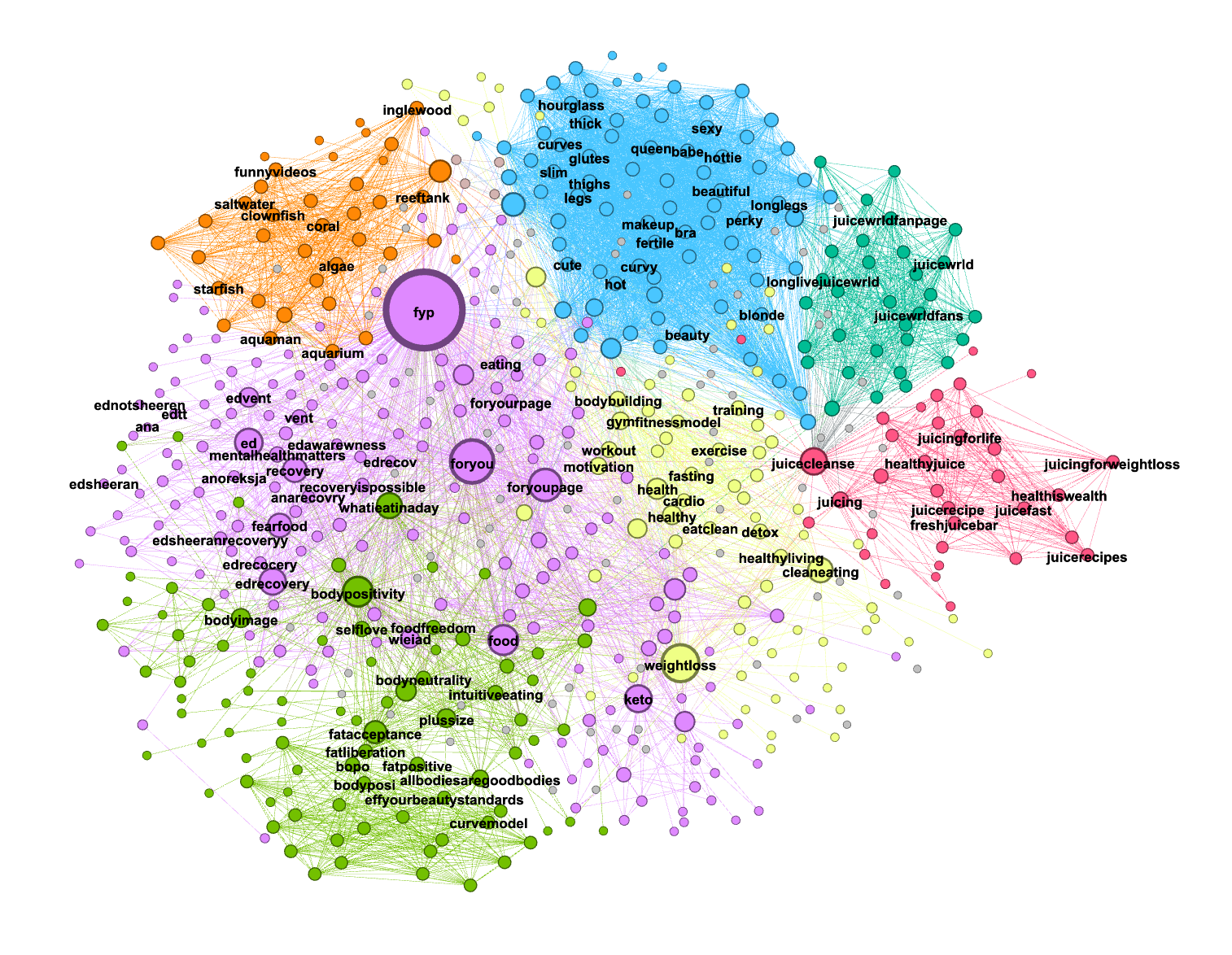}
    \caption{Purple = ED Recovery, Yellow = Healthy Living, Green = Body Positivity/Acceptance, Pink = Juice-Related, Blue = Beauty and Body Positivity, Teal = Music Promotion, Orange = Miscellaneous.}
    \label{fig:network}
\end{figure*}

\subsection{RQ1: Co-Occurrence of  Moderation Evaders and Mainstream Hashtags }
To address our first research question, we created a hashtag co-occurrence graph of popular hashtags  (that occurred at least 30 times), with 612 hashtags in total. Figure~\ref{fig:network} shows the hashtag co-occurrence network. The node size in this graph corresponds to the hashtag's PageRank centrality. We used the Louvain algorithm to identify clusters of highly interlinked nodes, revealing seven main communities. The central hashtags within these clusters are listed below, including those that are intentionally misspelled. 
\begin{itemize}
    \item Eating disorders (ED) recovery community with hashtags  \#edrecovery, \#edrecvery (\textit{sic}), \#anorexiarecovery, $\ldots$
    \item Healthy living community: \#exercise, \#workout, \#healthyrecipes, $\ldots$
    \item Body positivity/acceptance community:  \#bodypositivity, \#fatacceptance, \#selflove, $\ldots$
    \item Juice-related content: \#juicerecipes, \#healthyjuice, \#juicingforhealth, $\ldots$
    \item Beauty and body positivity community: \#thighs, \#beautiful, \#curvy, $\ldots$
    \item Music promotion: JuiceWRLD-related content community which include \#juicewrld \#juicewrldmusic, $\ldots$
    \item Miscellaneous: \#water \#aquarium
\end{itemize}

The largest cluster in this network, with 182 hashtags and over 1.9 million views per video, is devoted to \textit{ED recovery}.
Table~\ref{tab:averaged_modularity_stats} gives a full breakdown of the statistics for each cluster.

  \begin{figure*}[t]
    \includegraphics[width=\linewidth]{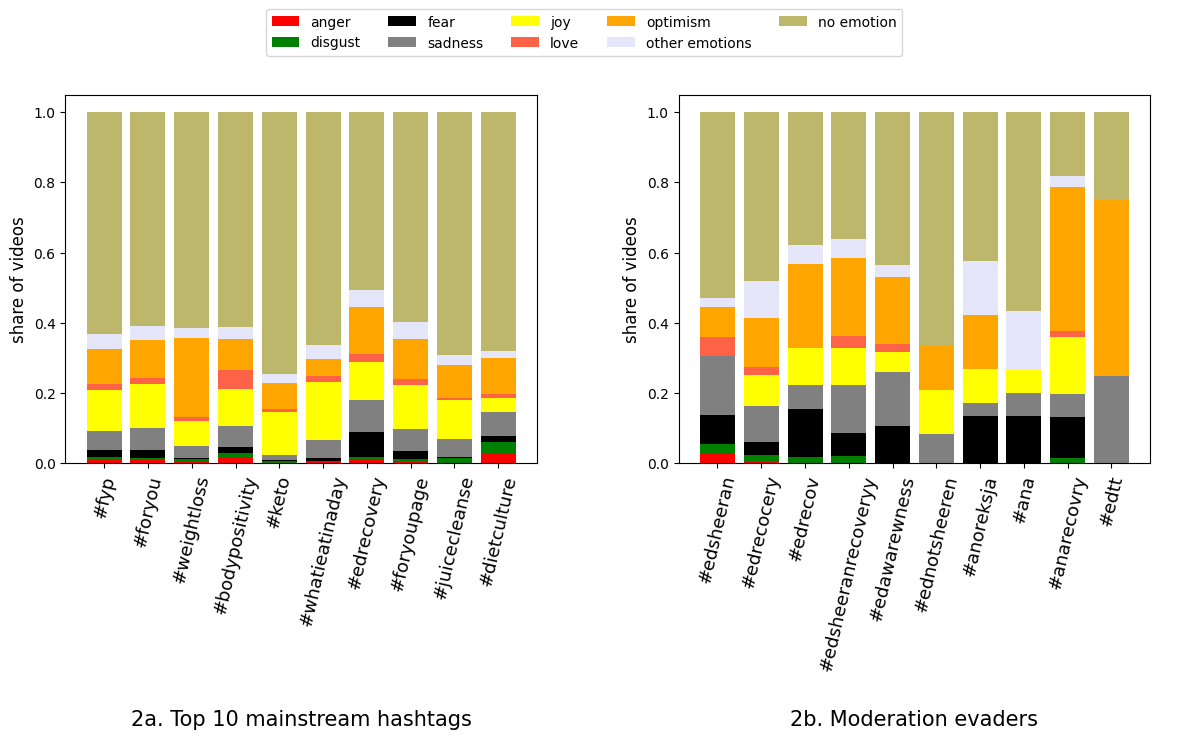}
    \caption{Comparative emotion analysis of video descriptions}
    \label{fig:emotionalanalysismainstream}
\end{figure*}


 We identified 10 moderation evaders within the 612 hashtags, and all of them were located in the ED recovery community. These hashtags were: \#edrecocery, \#edsheeranrecoveryy, \#edawarewness, \#anarecvery, \#edrecov, \#anoreksja, \#edtt, \#edsheeran, \#ednotsheeren, and \#ana. Moderation evaders frequently co-occurred with mainstream hashtags like \#fyp, \#fearfood, \#recovery, \#ed, and \#mentalhealthmatters. For instance, \#edrecocery appeared 138 times in total, with 64 occurrences alongside \#ed, while \#anarecovry appeared 61 times in total, with 55 occurrences with \#recovery. See table \ref{tab:cooccurrences} for full list.
 The co-occurrence of mainstream hashtags with moderation evaders suggests that users who search for terms such as \#ed or \#recovery could be exposed to content containing moderation evaders. 
 The hashtag \#anoreksja also co-occurred with \#tw (trigger warning) over 20 times. Furthermore, the hashtag \#wl (weight loss) co-occurred with \#edtt and \#ednotsheeren over 10 times. The associations of moderation evaders with \textit{weight loss} (\#wl) and \textit{trigger warnings} (\#tw) indicate potential triggering content that may be harmful to individuals struggling with eating disorders.

\paragraph{Findings and Remarks:} Addressing RQ1, the analysis of TikTok hashtags revealed a prominent cluster related to ED recovery, with 182 hashtags with over 1.9 million views per video.  
This contrasts with Twitter, where harmful content promoting eating disorders is far more common~\cite{lerman2023radicalized}. While this shows that TikTok provides a more positive, pro-recovery platform, the presence of moderation evaders within this cluster suggests that harmful content may be intentially obscured to avoid moderation. Moderation evaders are associated with hashtags weight loss (\#wl) and trigger warnings (\#tw), highlighting links to potentially problematic content in the context of eating disorders. Moreover, the fact that these moderation evaders are 
linked to mainstream hashtags, such as \#fyp and \#recovery, suggests potential exposure to pro-eating disorder content for users searching for recovery-related content.

\subsection{RQ2: Emotions in Video Descriptions} 
To address our second research question, we compared emotional expressions in the descriptions of TikTok videos tagged with mainstream hashtags and moderation evaders. 
We first focused on videos that have been tagged with any of the ten most popular mainstream hashtags (based on the number of occurrences). Joy consistently emerged as the dominant emotion, followed closely by optimism. Among negative emotions, sadness was the most common. A substantial portion of posts expressed ``no emotion'', potentially influenced by the succinct and hashtag-centric nature of TikTok video descriptions. See Figure~\ref{fig:emotionalanalysismainstream}a for the emotional analysis results for the mainstream hashtags.

 \begin{figure*}[t]
    \includegraphics[width=\linewidth]{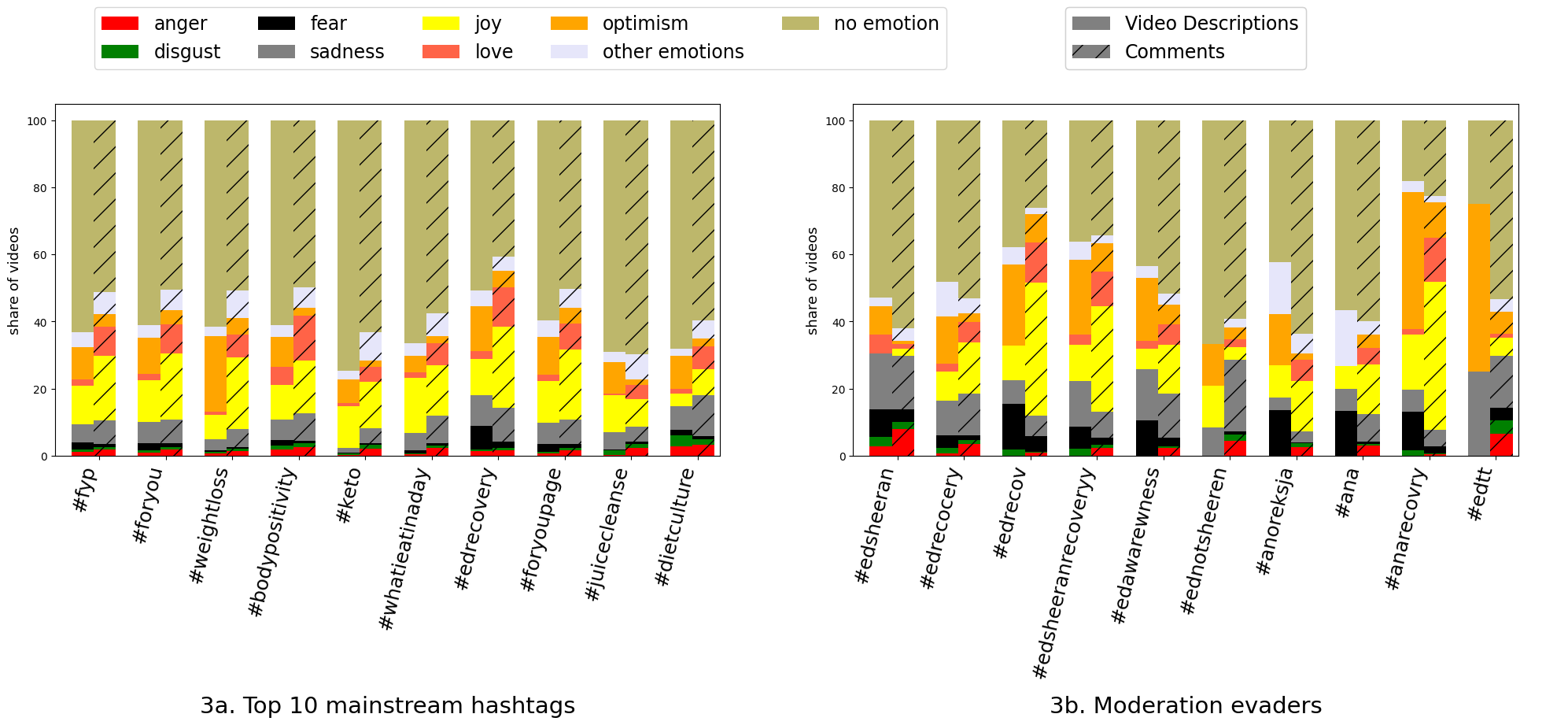}
    \caption{Comparative emotion analysis of video descriptions and comments}
    \label{fig:emotionalanalysismoderation}
\end{figure*}

Emotional analysis of the descriptions of videos tagged with any of the ten 
moderation evaders that we previously identified 
revealed them to be more emotional overall. 
Optimism was the dominant emotion. Sadness and fear were also higher, suggesting a more negative emotional tone (Figure~\ref{fig:emotionalanalysismainstream}b).

Comparing emotional expressions of videos tagged with mainstream hashtags and moderation evaders revealed interesting patterns. On the one hand, mainstream hashtags tended to be more positive overall, with higher occurrences of joy and love. Moderation evaders displayed a more diverse emotional landscape, with notable expressions of negative sentiments such as fear and sadness, underscoring the potential risks associated with these hashtags. The identified emotional differences have implications for user exposure. Mainstream content, characterized by positive emotions, may contribute to a more uplifting user experience. On the other hand, the diverse emotional landscape of moderation evaders' content, including expressions of fear and sadness, suggests potential exposure to content with a more complex emotional impact, indicating both positive and potentially harmful messaging. 
Figure \ref{fig:vidandcomments} lists sample video descriptions that express different emotions. 

To better understand these differences, we conducted a manual review of the video content for posts containing moderation evaders. The videos containing the five moderation evaders with the highest number of occurrences (\#edrecocery, \#edsheeranrecoveryy, \#edawarewness, \#anarecvery, and \#edrecov) often featured recovery content and positive messaging, providing tips on recovery approaches, highlighting the benefits of recovery, and addressing the toll of eating disorders.

Amongst the other five, the hashtags \#anoreksja, \#ednotsheeren, and \#ana were not searchable, i.e., they are blocked when searching for them in the TikTok search bar, and links to mental health resources are provided instead (see Figure~\ref{fig:anablocked} in the Appendix). It should be noted that the mainstream hashtag \#edrecovery was blocked as well.

Furthermore, an examination of posts containing \#edtt revealed that many of the videos included displays of eating disorders-related behaviors, dark humor, and an overall glorification of eating disorders.
Due to the co-occurrences of \#edtt with more mainstream hashtags such as \#edvent, \#ed, and \#fyp, this raises concerns about the potential exposure of users to harmful content of this nature. From the manual inspection of moderation evaders, we concluded they are not exclusively associated with harmful or negative content; instead, they encompass a variety of themes, including those focused on recovery and positive narratives related to eating disorders.

\paragraph{Findings and Remarks:}In response to RQ2, we explored the emotional expressions in TikTok video descriptions, revealing differences between mainstream hashtags and moderation evaders. Mainstream content tends toward positivity, with joy and optimism prevalent, while moderation evaders' content shows a wider emotional range, including fear and sadness. A manual review of moderation evaders underscores the need to understand TikTok's emotional dynamics and associated risks across various hashtags while also showing that TikTok has increased its moderation efforts by blocking some of these hashtags.

\subsection{RQ3: Emotions in Comments} 
Regarding RQ3, we investigated the relationship between emotional expressions in  TikTok video descriptions and the emotions expressed in the user comments in response to these videos. Our analysis revealed that comments displayed a broader range of emotions compared to video descriptions themselves, with an increase in anger (Fig.~\ref{fig:emotionalanalysismoderation}).

In general, the emotional analysis of the video descriptions and comments of mainstream hashtags showed a consistent pattern. In posts tagged with recovery-related hashtags,  both descriptions of videos and comments tended to be positive, with more expressions of joy, love, and optimism. This suggests that content associated with mainstream and recovery-related hashtags tends to foster positive engagement, creating a supportive and optimistic community atmosphere on TikTok. In comparison, the emotional analysis of video descriptions and comments associated with the moderation evaders reveals a clear emotional distinction. The user comments on these posts, while generally less emotionally charged than their corresponding video descriptions, still manifest a higher degree of joy, especially for the recovery-related hashtags.

Mainstream hashtags, while generally positive, exhibited slightly more anger in their comments. We also observed an increase in anger for the moderation evaders' posts comments in comparison to the video descriptions (see Fig.~\ref{fig:emotionalanalysismoderation}). The heightened expression of anger in comments, especially in moderation evaders' content, raises questions about the nature of discussions and interactions surrounding sensitive or controversial topics on TikTok. Figure \ref{fig:vidandcomments} lists sample comments that express different emotions. 

We have observed that some of the moderation evaders' comments displayed less emotion than the video descriptions, specifically \#edsheeran, \#edrecocery, \#edawarewness \#anoreskja, \#ana, \#anarecovry, and \#edtt. Also, the comments for the hashtags \#edsheeran, \#ednotsheeren, and \#edtt expressed less positive emotions compared to the video descriptions.

Interestingly, when analyzing the video descriptions for the hashtag \#ednotsheeren, it is evident that positive emotions such as joy and optimism are present but relatively subdued, with percentages ranging from 0\% to 12.5\%. Notably, the emotion love is not expressed in the video descriptions. On the contrary, negative emotions, specifically sadness, stand out at 8.33\%. Moving to comments, the disparity between positive and negative emotions becomes more pronounced. Positive emotions in comments, including joy, love, and optimism, collectively amount to 9.61\%, significantly lower than the corresponding video description percentage, which was 25\%. In contrast, negative emotions in comments, particularly sadness at 21.47\%, surpass the negative emotions in the video descriptions. This discrepancy suggests that the content under \#ednotsheeren may not be resonating positively with the audience, as indicated by the diminished expression of positive emotions in both video descriptions and comments.

Also - for the hashtag \#edtt - even though the video descriptions contained mostly positive emotions (with 50\% being labeled with optimism), this did not mean that the viewers exhibited the same sentiment. For the comments, the negative emotions total 29.86\% while the positive emotions come to 12.98\%. Despite 50\% of the posts being classified as \textit{optimism}, it appears that viewers might not perceive the content in the same way. This suggests that there may be a gap between the positive message in the video descriptions and how the audience actually interprets it.





\paragraph{Findings and Remarks:} For RQ3, we examined the relationship between emotional expressions in video descriptions and comments. We observed that comments on videos with mainstream hashtags exhibited more emotions than video descriptions themselves, while moderation evaders generally elicited fewer emotions. Furthermore, there was a noticeable increase in negative emotions expressed in comments for moderation evaders compared to mainstream hashtags. This hints that some moderation evaders could hide potentially problematic content.


\section{Discussion}

We detected various TikTok communities related to eating disorders recovery, body positivity, and healthy living. These communities mostly contain content that is designed to be informational, inspirational, or uplifting in some shape or form; however, it should be noted that some of this well-intentioned content can have adverse effects as well. For example, content showing a person's body can encourage negative comparison and lead to viewers feeling poorly about their own bodies. Additionally, weight loss tips from unqualified TikTok users may be harmful to people struggling with eating disorders.
Prior work done by the CCDH\textsuperscript{\ref{note1}}
and \cite{lerman2023radicalized} demonstrated that users within the ED community utilize misspellings, abbreviations, and the musical artist Ed Sheeran's name to avoid moderation. Within our dataset, we found 10 hashtags fitting this criterion, with each appearing in over 30 posts, all within the ED recovery/positivity community. These hashtags were found to co-occur with mainstream hashtags such as \#fearfood, \#mentalhealthmatters, and \#recovery. This, combined with the fact that 1) the video descriptions with moderation evaders tend to evoke more negative emotions and 2) videos with some moderation evaders contain content that could promote ED behaviors, suggests that vulnerable users could potentially stumble upon harmful content even if they search for mainstream topics. 

Nevertheless, the fact that some of these hashtags are blocked from the search page shows that TikTok has likely made an effort to moderate content that could potentially promote eating disorder behaviors, which may explain the overall positive tone of the content. The hashtag \#edtt was one that was not blocked, though, and appeared to contain harmful content with ``dark humor", ED-glorification, and ED-promoting advice. This is a hashtag that should be investigated further in future studies that analyze a larger amount of video content. 

The emotional analysis for the video descriptions of the mainstream hashtags showed that they generally displayed a more balanced emotional distribution, representing a wide array of positive and neutral emotions. This could be because mainstream hashtags are often more visible to a broader audience which may lead users to adopt a more restrained and balanced emotional tone to appeal to a diverse viewership. On the other hand, moderation evaders' video descriptions tended to be more emotionally charged, with an increased presence of sadness, fear, and anger. This could suggest that those trying to avoid moderation are deliberately posting more negative content, believing that their content is less likely to be taken down.

The user comments on posts containing mainstream hashtags were more emotionally charged than their respective video descriptions. This could possibly be because users tend to be more careful and deliberate when sharing videos under popular hashtags, aiming to present a carefully curated image to a larger audience.
These comments revealed a similar pattern to that of the video descriptions -- joy and optimism were often the most common emotions. In contrast, the user comments on posts containing the moderation evaders were less emotionally charged than their respective video descriptions. Despite being less emotionally charged overall, these comments contained a higher amount of joy compared to their video descriptions, especially for hashtags related to recovery. This could indicate potential community formation where individuals show support, particularly concerning eating disorder recovery. Amongst both the comments for the mainstream hashtags and moderation evaders, anger was more present also - albeit still at a minimal level. 

The discovery of moderation evaders, their co-occurrences with more mainstream hashtags, and the persistence of harmful content despite some blocked hashtags underscore the ongoing challenges in content moderation. The emotional analysis indicates that posts containing mainstream hashtags tend to display a balanced emotional distribution in the video description and comments, possibly due to broader visibility, whereas moderation evader hashtags exhibit more emotionally charged content, emphasizing the need for targeted moderation efforts to mitigate negative emotional impacts.

\subsection{Ethics Statement}
All data used for this study is public and collected following
TikTok’s terms of service. In our analysis of TikTok videos and comments within these videos, no identifiable information related to any user has been included and analysis was carried out on aggregated data. These steps ensure that negative outcomes due to use of these data are minimized.

The authors declare no competing interests.

\subsection{Limitations}
It is crucial to acknowledge the limitations of our analysis. The prevalence of ``no emotion'' in our analysis may be influenced by the hashtag-centric nature of TikTok video descriptions, potentially impacting the accuracy of emotional expression detection. Additionally, our emotional analysis was mostly confined to the textual video descriptions and comments, overlooking the visual and audio elements of the videos. While we manually reviewed some of the videos containing specific hashtags, this inspection was not comprehensive, and future research should look into analyzing these elements further while incorporating AI techniques and statistical analysis.


\subsection{Conclusion}
Our results provide valuable insights into the emotional dynamics of TikTok content concerning mental health and eating disorders -- specifically content containing moderation evader hashtags. We have found that moderation evaders do co-occur with mainstream hashtags and that the video descriptions/comments of posts containing the former tend to be more emotionally charged (including a higher amount of negative emotions such as fear and sadness). The findings have implications for content creators, platform moderation, and interventions aimed at fostering a supportive online environment for discussions on mental health and eating disorders. Future studies should continue to inspect moderation evaders and add onto our findings by taking a comprehensive look at the video content of posts in addition to their video descriptions and comments.

{
\bibliography{aaai22}
}
\newpage
\section{Appendix}
\appendix

\textbf{Search Terms}
thinspo, proana, proanatips, anatips, meanspo, fearfood, sweetspo, eatingdisorder, bonespo, promia, redbracetpro, m34nspo, fatspo, lowcalrestriction, edvent, WhatIEatInADay, Iwillbeskinny, thinspoa, ketodiet, skinnycheck, thighgapworkout, bodyimage, bodygoals, weightloss, skinnydiet, chloetingchallange, fatacceptance, midriff, foodistheenemy, cleanvegan, keto, cleaneating, intermittentfasting, juicecleanse, watercleanse, EDrecovery, bodypositivity, dietculture.

\begin{table}[h]
    \centering
    \caption{Co-occurrences of Moderation Evader Hashtags with Some Mainstream Hashtags}
    \begin{adjustbox}{width=\columnwidth}
        \begin{tabular}{|l|c|c|c|c|}
            \hline
            Hashtag & Total Occurrences & \#recovery & \#ed & \#edvent\\
            \hline
            \#edrecocery & 138 & 69 & 64 & 21\\
            \#edsheeranrecoveryy & 102 & 57 & 31 & 16\\
            \#edawarewness & 86 & 43 & 38 & 37\\
            \#anarecovry & 61 & 55 & $<10$ & $<10$\\
            \#edrecov & 58 & 44 & 22 & 16\\
            \#anoreksja & 47 & 38 & 37 & $<10$\\
            \#edtt & 39 & $<10$ & 35 & 39\\
            \#edsheeran & 38 & 15 & $<10$ & 14\\
            \#ednotsheeren & 36 & $<10$ & 22 & 31\\
            \#ana & 31 & 16 & 25 & $<10$\\
            \hline
        \end{tabular}
    \end{adjustbox}
    \label{tab:cooccurrences}
\end{table}

 \begin{figure}[h]
    \centering
    \includegraphics[width=0.5\columnwidth]{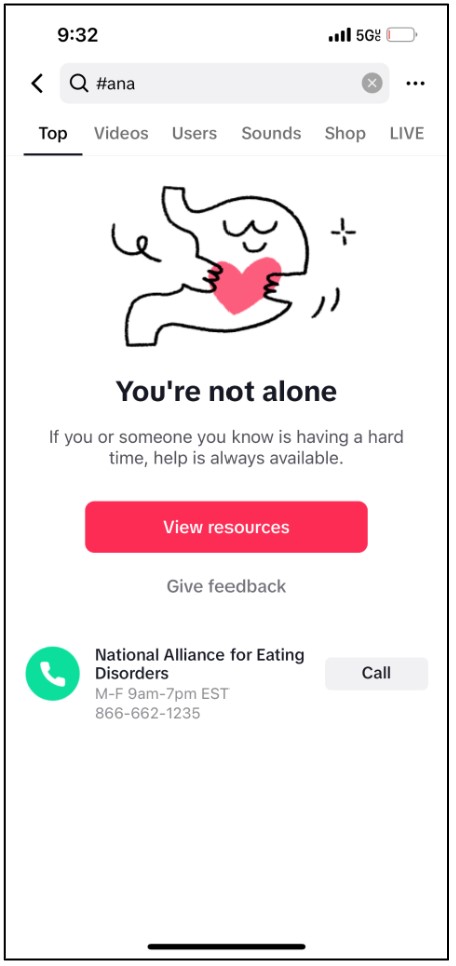}
    \caption{Hashtag \#ana blocked on TikTok, redirecting users to mental health resources}
    \label{fig:anablocked}
\end{figure}

\newpage
\begin{figure}
    \centering
    \begin{tikzpicture}
        \node[fill=mygray, rounded corners, drop shadow={fill=black!30, shadow xshift=3pt, shadow yshift=-3pt, opacity=0.5}, inner sep=10pt] {
            \begin{minipage}{0.5\textwidth}
                \begin{tabular}{l p{0.5\textwidth}}
                    \texttt{Emotion:} & 
                       Joy\\
                      
                    \texttt{Video Description:} & I did it! Let’s get the big girls to go viral \#fy \#fat \#big \#bbw \#curvy \#plussize \#biggirl \#bodypositivity \#fatacceptance \#stitch \#ArmaniMyWay \\
                     \texttt{Comment:} & Think about how sweet and amazing it tastes remember the feeling of accomplishment once you have finished eating xx Stay strong\\
                     \\
                    \texttt{Emotion:} & Optimism \\
                      \texttt{Video Description:} & so much work to do but we gon’ do it! \#dietculture \#bodyimage \#caloriedeficit \#selflove \#toxicrelationship \#edrecovery\\
                      \texttt{Comment:} & Im going to watch this everyday until i reach my goal this is real motivations right here\\
                      \\
                      \texttt{Emotion:} & Sadness \\
                      \texttt{Video Description:} & Spent a long time wishing things like this would happen, just to shrink my body :( \#PlusSize \#FatLiberation \#FatAcceptance \#BodyPositivity \#FatTikTok \#ThatFatBaddee  \#fatacceptance \#stitch \#ArmaniMyWay\\
                      \texttt{Comment:} &  I started crying when it said “Go help other girls!”\\
                      \\
                      \texttt{Emotion:} & Fear \\
                      \texttt{Video Description:} & Not knowing the exact amount was very scary, but I'm so sick of this scale!! \#ed \#recovery \#eatittobeatit \#edrecovery \#edrecocery \#foodisfuel \#fearfood \#fearfoodchallenge \#fyp \#fy\\
                      \texttt{Comment:} & As a nutritionist this is TERRIFYING. Whyyyyy would they even publish this?\\
                \end{tabular}
            \end{minipage}
        };
    \end{tikzpicture}
    \caption{Samples of video descriptions and comments expressing emotions}
    \label{fig:vidandcomments}
\end{figure}

\end{document}